\documentstyle[preprint,aps]{revtex}

\begin{document}
\title{Theory of the Quantum Speed Up}
\author{Giuseppe Castagnoli,}
\address{Elsag, 16154 Genova, Italy}
\author{David Ritz Finkelstein,}
\address{Georgia Institute of Technology, Atlanta, USA}
\date{\today}
\maketitle

\begin{abstract}
{Insofar as quantum computation is faster than classical, it appears to be
irreversible. In all quantum algorithms found so far the speed-up depends on
the extra-dynamical irreversible projection representing quantum
measurement. Quantum measurement performs a computation that dynamical
computation cannot accomplish as efficiently.}
\end{abstract}

\section{Premise}

\noindent The quantum algorithms are sometimes faster than their classical
counterparts. We show that this quantum speed-up results from a succession
of entanglement and disentanglement, the former due to dynamical
quantum-parallel computation, the latter to the extra-dynamical projection
of quantum measurement. Thus the quantum speed-up implies irreversibility.

Some standard notions concerning problem solving must be modified to
understand the speed-up. Standard problem solving has three stages:

\begin{itemize}
\item[(i)]  {\em State the problem}. This defines the problem solution,
usually implicitly. E.g. consider the problem of finding two primes $x,y$
(unknown) such that $x\cdot y=c$ (known). This equation implicitly defines
the values of $x$ and $y$ which satisfy it. An implicit definition does not
represent the process required to compute the solution.

\item[(ii)]  {\em Program the computation}. Change the implicit definition
into an explicit, finite step-by-step logical procedure for constructing the
solution. This procedure is specified by the solution algorithm.

\item[(iii)]  {\em Run the program}. The execution is dynamical in
character. By dynamics we mean, here and in the following, deterministic
dynamics\footnote{%
Classical nondeterministic computation, at the current fundamental level,
will be seen as pseudorandom deterministic computation.}.
\end{itemize}

\noindent The standard assumption is that the solution of a problem must be
computed by a dynamical development. Step (ii) changes a definition which
does not represent a dynamical process, into one which represents it.

Quantum computation does not fit this scheme, as we will see in the next
Section.

\section{The speed-up in Shor's algorithm}

\noindent Consider Shor's algorithm \cite{SHOR}. The problem is to
efficiently find the period $r$ of a hard-to-reverse function $f(x)$ from $%
\left\{ 0,1\right\} ^{n}$ to $\left\{ 0,1\right\} ^{n}$. Fig. 1 gives the
algorithm block diagram.

\begin{center}
Fig. 1
\end{center}

\noindent $H$ is the Hadamard and $F$ the digital Fourier transform, $M$
denotes measurement of a register content. We need to consider only two
steps of the algorithm (see Castagnoli et al., \cite{CASTETAL}).

\begin{itemize}
\item[(I)]  The process of computing $f(x)$ for all possible $x$, in quantum
superposition, puts two n-qubit registers $X$ and $F$ into the state 
\footnote{%
Quantum theory can be formulated praxically or ontically. The former (for
example, Finkelstein \cite{FINK96}) is closest to Heisenberg's. It deals
with operators dispensing with states. The ontic formulation makes quantum
theory seem less time-symmetrical than it really is. In the present theory,
it makes the speed-up seem to happen all at once at the end of computation.
We use the ontic formulation here, misleading as it is, because it is more
familiar to most physicists.}
\end{itemize}

\[
\left| \psi ,t_{2}\right\rangle _{XF}=\frac{1}{\sqrt{N}}\sum_{x}\left|
x\right\rangle _{X}\left| f\left( x\right) \right\rangle _{F}, 
\]

$x\;$runs over $0,1,...,N-1,$ with $N=2^{n}.$

\begin{itemize}
\item[(II)]  Let $\left[ F\right] $ be the content of register $F$, an
observable. Measuring $\left[ F\right] $ in $\left| \psi ,t_{2}\right\rangle
_{XF}$ \footnote{%
This intermediate measurement can be skipped, but we will show that this
makes no difference.} and finding the result $\overline{f}$ yields the state
\end{itemize}

\[
\left| \psi ,t_{3}\right\rangle _{XF}=k\left( \left| \overline{x}%
\right\rangle _{X}+\left| \overline{x}+r\right\rangle _{X}+\left| \overline{x%
}+2r\right\rangle _{X}+...\right) \left| \overline{f}\right\rangle _{F}; 
\]

$k$ is for normalization, $f(\overline{x})=f(\overline{x}+r)=...=\overline{f}%
.$

The transition from $\left| \psi ,t_{2}\right\rangle _{XF}$ to $\left| \psi
,t_{3}\right\rangle _{XF}$ obeys the quantum principle:

\begin{itemize}
\item[(A)]  measuring an observable is extra-dynamically represented by
projection on the eigenspace of one eigenvalue;

\item[(B)]  this eigenvalue is selected at random according to the square of
the probability amplitudes.
\end{itemize}

\noindent If $Q$ and $Q^{^{\prime }}$ are the projection operators on the
state just before and after the measurement $M$, and $P$ is the projection
operator for the observed value of the function, then classically $%
Q^{^{\prime }}=Q$, while quantally $Q^{^{\prime }}=PQP$ (up to
normalization), depending on both $P$ and $Q$.

Thus, selecting $\overline{f}$ projects into the post-measurement state all
and only those tensor products of $\left| \psi ,t_{2}\right\rangle _{XF}$
ending with $\left| \overline{f}\right\rangle _{F}$. Point (A) of the
quantum principle, by selecting one eigenvalue, imposes a logical constraint
on the output of computation. Because of this constraint, quantum
measurement filters, out of an exponentially larger superposition, all and
only those values of $x$ whose function is that eigenvalue.

Therefore quantum measurement performs extra-dynamically a computation
crucial for finding $r$, which is ``readily'' extracted out of $\left| \psi
,t_{3}\right\rangle _{XF}$. Measurement time is linear in the number of
qubits of register $F$, and is independent of the entanglement between $X$
and $F$, which holds problem complexity. Disentanglement comes for free, as
a by-product of quantum measurement.

Filtration, together with function evaluation, is essential to speed-up.
This can be better seen by comparing step by step quantum and classical
computation times\footnote{%
Classical time is that required to derive the {\em symbolic description} of
a quantum state (e.g. $\left| \psi ,t_{2}\right\rangle _{XF}$ , or $\left|
\psi ,t_{3}\right\rangle _{XF}$, with all $x$, $f\left( x\right) $, $%
\overline{x}$, $r$, etc. replaced by the proper numerical values) from the
previous one by classical computation. Note that the resulting classical
algorithm is reasonably efficient in itself, with times on the order of
problem size.}: (I) function evaluation: poly$\left( n\right) $ vs exp$%
\left( n\right) ;$ (II) filtration: linear$\left( n\right) $ vs exp$\left(
n\right) $; (III) extracting $r$ out of $\left| \psi ,t_{3}\right\rangle
_{XF}$: linear$\left( n\right) $ vs linear$\left( n\right) .$ Speed-up is
due to steps (I) and (II).

The extra-dynamical character of quantum computation is clarified by showing
that Shor's algorithm does not fit standard, dynamical, problem solving.
Quantum dynamics is deterministic: any state in time dynamically determines
a unique successor. While function evaluation is dynamical in character, the
filtration performed by quantum measurement is not.

Classically the state after measurement $M$ is the same as the state before $%
M$. In quantum fact the state after measurement is influenced by both the
state before measurement and the measurement itself, by the quantum
principle: $\left| \psi ,t_{3}\right\rangle _{XF}=k\left| \overline{f}%
\right\rangle _{F}\left\langle \overline{f}\right| _{F}\left| \psi
,t_{2}\right\rangle _{XF}.$ \noindent $\left| \psi ,t_{2}\right\rangle _{XF}$
is the {\em prior state}; the left-multiplication by $\left| \overline{f}%
\right\rangle _{F}\left\langle \overline{f}\right| _{F}$ represents the
final constraint selecting all tensor products ending with $\left| \overline{%
f}\right\rangle _{F}$. The determination of $\left| \psi ,t_{3}\right\rangle
_{XF}$ is jointly influenced by an initial condition and a final condition.
It is richer than dynamical determination, insofar as it yields the
speed-up. We needed a computational context to realize this.

Extra-dynamical computation means much more than nondeterministic
computation. For example, point (A) of the quantum principle does not
involve randomness and yields Shor's quantum speed-up in $\sim70\%$ of the
cases, when a single run of the algorithm is sufficient to identify $r$ 
\footnote{%
In $\sim30\%$ of the cases, more than one run is needed. Randomness assures
that we do not always obtain the same result.}.

Determination with joint influence is extra-dynamical, it cannot be
represented by a dynamical propagation of an input into an output. Of course
we could go through step (ii), and replace joint influence with a dynamical
process that leads to a ``filtered'' state like $\left| \psi
,t_{3}\right\rangle _{XF}$. But this would introduce programming and
computation, increasing computation time exponentially in problem size.
Joint influence bypasses step (ii) as well as speeding-up the computation.
It yields a direct physical determination of the object of an implicit
definition\footnote{%
Since the implicit definition is the problem, we see that, in Shor's
algorithm, computation can be identified with problem solving: points (i),
(ii), and (iii) of Section I are both altered and unified.}.

This can also be seen as follows. $\left| \psi,t_{3}\right\rangle _{XF}$
(selected with joint influence) contains the solutions $x$ of the implicit
algebraic equation $f(x)=\overline{f}$, although the reverse of $f$ has not
been computed. Thus, $f(x)=\overline{f}$ implicitly defines the solutions $x$
while quantum measurement selects them without going through programming and
dynamical computation. The extra-dynamical character of this selection is
clear. It takes essentially no time.

It can be seen that the same theory of the speed-up holds for Simon's
algorithm \cite{SIMON94}, as modified in \cite{CLEVE96}.

Until now we have assumed the intermediate measurement of $\left[ F\right] $%
. However, as is well known, this measurement can be skipped without
affecting the result of measuring $\left[ X\right] $ at time $t_{4}$ (fig.
1). It was introduced by Ekert and Jozsa \cite{EK-JOZ} to clarify the way
Shor's algorithm operates; it can also clarify the speed-up. In fact
skipping it is mathematically{\em \ }equivalent to performing it: the
filtration performed by the extra-dynamical projection of quantum
measurement is induced by measuring only $\left[ X\right] $ at the end.

If $\left[ F\right] $ measurement is skipped, the state of registers $X$ and 
$F$ at time $t_{4}$ is entangled. This establishes an equivalence between
measuring $\left[ X\right] $ or $\left[ F\right] $. From a mathematical
standpoint, the outcome of measuring $\left[ F\right] $ at time $t_{4}$ can
be backdated in time along the reversible process, provided that the overall
state undergoes the inverse of the usual forward-time evolution. This is
equivalent to having measured $\left[ F\right] $ at time $t_{2}$.

We should counter the objection that register $F$ can be annihilated
immediately after function evaluation. This would leave register $X$ in a
mixture that is the partial trace over $F$ of the density matrix of the two
registers: 
\[
\left| \psi ,t_{2}\right\rangle _{X}^{^{\prime }}=\frac{1}{\sqrt{N}}%
\sum_{h}e^{i\delta _{h}}\left( \left| x_{h}\right\rangle _{X}+\left|
x_{h}+r\right\rangle _{X}+\left| x_{h}+2r\right\rangle _{X}+...\right)
\left| f_{h}\right\rangle _{F}, 
\]
where the range of $h$ is such that $f_{h}$ ranges over all the values
assumed by $f\left( x\right) $, $f\left( x_{h}\right) =f\left(
x_{h}+r\right) =...=f_{h}$, and $\delta _{h}$ are random phases independent
of each other\footnote{%
We are using the random phase representation. Let us exemplify it for a
two-state system. The mixture $\rho =\sin ^{2}\varphi \left| 0\right\rangle
\left\langle 0\right| +\cos ^{2}\varphi \left| 1\right\rangle \left\langle
1\right| $ becomes $\left| \psi \right\rangle =\sin \varphi \left|
0\right\rangle +e^{i\delta }\cos \varphi \left| 1\right\rangle $, where $%
\delta $ is a random phase with uniform distribution in $\left[ 0,2\pi %
\right] $; $\rho $ is the average over $\delta $ of $\left| \psi
\right\rangle \left\langle \psi \right| $.}. For the current purposes,
annihilating $F$ is like having performed the intermediate $\left[ F\right] $
measurement.

\section{The speed-up in quantum oracle computing}

\noindent Quantum oracle computing can be seen as a competition between two
players. One produces the problem, the other is challenged to produce the
solution. We shall call the former player Sphinx, the latter Oedipus.

Let us consider Grover's algorithm \cite{GROVER}. The game is as follows.
The Sphinx hides an object in drawer number $k$, among $n$ drawers. Oedipus
must find where it is, in the most efficient way. The chest of drawers is
actually a quantum computer that, set in the mode $k$ and given a drawer
number $x$ as the input, yields the output $f_{k}\left( x\right) =\delta
_{k,x}\ $($\delta _{k,x}=1$ if $k=x$ and $\delta _{k,x}=0$ if $k\neq x$).
Fig. 2a gives the usual Grover's algorithm for $n=4$.

\begin{center}
Fig. 2a,b
\end{center}

The Sphinx sets the mode $k$ at random and passes the computer on to
Oedipus. Oedipus must find $k$ in the most efficient way by testing the
computer input-output behaviour.

Without entering into detail, we note that the computer has two registers $X$
and $F$. Oedipus prepares them in the initial state $\frac{1}{\sqrt{2}}%
\left| 00\right\rangle _{X}\left( \left| 0\right\rangle _{F}-\left|
1\right\rangle _{F}\right) ,$ \noindent the same for all $k$, and invokes
the algorithm presented in fig. 2a. \noindent The state $\psi $ before final
measurement depends on the Sphinx' choice $k$:

\begin{eqnarray*}
k=00 &\leftrightarrow &\psi =\frac{1}{\sqrt{2}}\left| 00\right\rangle
_{X}\left( \left| 0\right\rangle _{F}-\left| 1\right\rangle _{F}\right) , \\
k=01 &\leftrightarrow &\psi =\frac{1}{\sqrt{2}}\left| 01\right\rangle
_{X}\left( \left| 0\right\rangle _{F}-\left| 1\right\rangle _{F}\right) , \\
k=10 &\leftrightarrow &\psi =\frac{1}{\sqrt{2}}\left| 10\right\rangle
_{X}\left( \left| 0\right\rangle _{F}-\left| 1\right\rangle _{F}\right) , \\
k=11 &\leftrightarrow &\psi =\frac{1}{\sqrt{2}}\left| 11\right\rangle
_{X}\left( \left| 0\right\rangle _{F}-\left| 1\right\rangle _{F}\right) .
\end{eqnarray*}

\medskip

Measuring $\left[ X\right] $ yields Oedipus answer. This is reached in $%
O\left( \sqrt{n}\right) $ time, versus $O\left( n\right) $ with classical
computation. But it is reached in a dynamical way, without any interplay
between quantum parallel computation and the extra-dynamical projection of
quantum measurement.

As we have seen, this interplay is associated with an isomorphism between
the problem that implicitly defines its solution and the solution
determination. This obviously requires that the problem is physically
represented in a complete way. Here it is not:\ the above possible choices
of the Sphinx and the related implications are not physically represented.

This is easily altered by introducing an ancillary two-qubit register $K$
which contains the computer mode $k$. Given the input $k$ and $x$, the
output of computation is now $F\left( k,x\right) =f_{k}\left( x\right) $
(fig. 2b). The preparation becomes

\[
\frac{1}{2\sqrt{2}}\left( \left| 00\right\rangle _{K}+e^{i\delta _{1}}\left|
01\right\rangle _{K}+e^{i\delta _{2}}\left| 10\right\rangle _{K}+e^{i\delta
_{3}}\left| 11\right\rangle _{K}\right) \left| 00\right\rangle _{X}\left(
\left| 0\right\rangle _{F}-\left| 1\right\rangle _{F}\right) , 
\]

\noindent where $\delta _{1}$, $\delta _{2}$ and $\delta _{3}$ are
independent random phases. To Oedipus, the Sphinx' random choice is
indistinguishable from a mixture where $k$ is a random variable with uniform
distribution over $00$, $01$, $10$, $11$. Fig. 2b includes the physical
representation of the problem; we can go directly to the final state before
measurement:

\[
\left( \left| 00\right\rangle _{K}\left| 00\right\rangle _{X}+e^{i\delta
_{1}}\left| 01\right\rangle _{K}\left| 01\right\rangle _{X}+e^{i\delta
_{2}}\left| 10\right\rangle _{K}\left| 10\right\rangle _{X}+e^{i\delta
_{3}}\left| 11\right\rangle _{K}\left| 11\right\rangle _{X}\right) \left(
\left| 0\right\rangle _{F}-\left| 1\right\rangle _{F}\right) , 
\]

\noindent $\delta _{1}$, $\delta _{2}$ and $\delta _{3}$ are independent
random phases. Measuring $\left[ K\right] $ gives the Sphinx' choice,
measuring $\left[ X\right] $ gives Oedipus answer, or vice-versa. Now that
the game has been physically represented, we find again that there is the
above ``interplay''.

In this game context, joint influence becomes the joint determination of the
drawer number on the part of the two players, imposed by the quantum
principle.

Why does extra-dynamical joint influence produce a speed-up? We suggest the
following argument. The quantum game -- yielding joint determination of the
drawer number as a special quantum feature -- should be as efficient as a
classical game where there were joint determination of the drawer number on
the part of the two players.

This cannot mean that Oedipus dictates the Sphinx' choice, or the Sphinx
suggests to Oedipus the right answer: this would be unilateral
determination. Joint determination is symmetrical. The classical game must
be defined as follows, with reference to the square-shaped chest of drawers
herebelow\footnote{%
If the number of drawers is $O\left( n\right) $, the number of rows or
colums is $O\left( \sqrt{n}\right) $.}

\begin{center}
\begin{tabular}{ccc}
& $0$ & $1$ \\ \cline{2-3}
$0$ & \multicolumn{1}{|c}{$00$} & \multicolumn{1}{|c|}{$01$} \\ \cline{2-3}
$1$ & \multicolumn{1}{|c}{$10$} & \multicolumn{1}{|c|}{$11$} \\ \cline{2-3}
\end{tabular}

\medskip
\end{center}

The Sphinx chooses the row number, say 1. Oedipus chooses the column number,
say 0. Clearly, the drawer number 10 has been jointly determined by the
Sphinx and Oedipus. Now the cost of Oedipus search is $O(\sqrt{n})$ rather
than $O(n)$, since he must search only the row. This is in agreement with
theory.

A similar analysis applies to Deutsch's algorithm \cite{DEUTSCH} as modified
in \cite{CLEVE96}.

\section{Conclusions}

We have shown that quantum computation speed-up depends essentially on the
extra-dynamical, irreversible projection of quantum measurement. To be sure,
the entropy increase associated with speed-up is proportional only to the
size of the output register, not the computation.

Extra-dynamical computation is more efficient than dynamical computation, as
it yields the speed-up. It is a high level quantum feature, as it comes from
a special interplay between a plurality of lower level ones (entanglement,
disentanglement ...).

Earlier, attention was paid only to reversible quantum computation. The
seminal well known works of Bennett, Fredkin and Toffoli, Benioff, and
Feynmann demonstrated that computation can be reversible both in the
classical and quantum framework. With Deutsch and others, quantum
computation becomes quantum problem-solving, yields a speed-up and, we point
out, ceases to be reversible. The current quantum algorithms ingeniously
exploit extra-dynamical computation.

It is natural to ask whether other extra-dynamical projections than the one
inherent in quantum measurement can be useful. Two come to mind at once: a
statistics symmetry can be seen as an extra-dynamical projection on the
``symmetric'' subspace; and annealing is a projection on the ground state
resulting from gradual cooling by a succession of extra-dynamical
interventions. Exploiting these forms of projection might result in further
speed-ups, and further reductions in the programming process.

This work developed through many discussions with Artur Ekert.

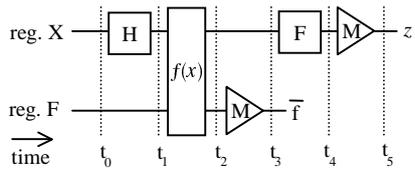
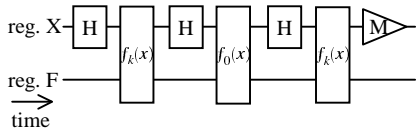
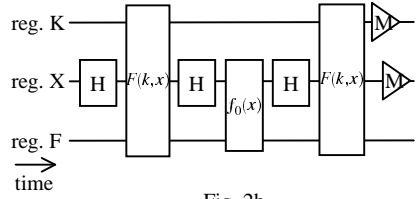
\end{document}